\documentclass[final]{aa}
\usepackage{graphicx}
\usepackage{natbib}

\newcommand{\etal}{et al.\ }
\newcommand{\msun}{{\,\rm M}_{\odot}}
\newcommand{\lsun}{{\,\rm L}_{\odot}}

\newcommand{\nht}{\ifmmode {{\rm NH}_3} \else {NH{\bas 3}} \fi}
\newcommand{\tco}{\ifmmode {^{13}{\rm CO} } \else {$^{13}{\rm CO}$ }\fi}
\newcommand{\dco}{\ifmmode {^{12}{\rm CO} } \else {$^{12}{\rm CO}$ }\fi}
\newcommand{\cdo}{\ifmmode {{\rm C}^{18}{\rm O} } \else {${\rm C}^{18}{\rm O}$ }\fi}

\newcommand{\htco}{\ifmmode {{\rm H}^{13}{\rm CO}^{+} } \else {${\rm H}^{13}{\rm CO}^{+}$ }\fi}
\newcommand{\hco}{\ifmmode {{\rm H}^{12}{\rm CO}^{+} } \else {${\rm H}^{12}
{\rm CO}^{+}$ }\fi}
\newcommand{\juz}{\ifmmode {{\rm J}=1\!\rightarrow\!0} \else
{${\rm J}=1\!\rightarrow\!0$} \fi}
\newcommand{\jdu}{\ifmmode {{\rm J}=2\!\rightarrow\!1} \else
{${\rm J}=2\!\rightarrow\!1$} \fi}
\newcommand{\jtd}{\ifmmode {{\rm J}=3\!\rightarrow\!2} \else
{${\rm J}=3\!\rightarrow\!2$} \fi}
\newcommand{\jcq}{\ifmmode {{\rm J}=5\!\rightarrow\!4} \else
{${\rm J}=5\!\rightarrow\!4$} \fi}
\newcommand{\as}{\ifmmode {^{\scriptscriptstyle\prime\prime}}
\else $^{\scriptscriptstyle\prime\prime}$\fi}
\newcommand{\am}{\ifmmode {^{\scriptscriptstyle\prime}}
\else $^{\scriptscriptstyle\prime}$\fi}
\begin{document}
 
\title{A Keplerian disk around the Herbig Ae star HD 34282}

\subtitle{}

\author{V. Pi\'etu \and A.Dutrey  \and C. Kahane}

\offprints{Vincent.Pietu@obs.ujf-grenoble.fr}

\institute{Laboratoire d'AstrOphysique de Grenoble, B.P.53, F-38041 Grenoble
Cedex 9, France \\ \email{Vincent.Pietu@obs.ujf-grenoble.fr}}

\date{Received 9 July 2002 ; accepted 18 September 2002}


\abstract{
We report new millimeter observations of the circumstellar material surrounding the Herbig Ae A0.5 star HD 34282 performed with the IRAM array in CO~\jdu~ and in continuum at 1.3 mm. These observations have revealed the existence of a large Keplerian disk around the star. We have analysed simultaneously  the line and   continuum emissions to derive the physical properties of both the gas and the dust. The analysis of our observations also shows that the Hipparcos distance to the star is somewhat underestimated ; the actual distance is probably about 400 pc. With this distance the disk around HD 34282 appears more massive and somewhat hotter than the observed disks around less massive T Tauri stars, but shares the general behaviour of passive disks. 
}

\maketitle


\section{Introduction}

Interferometric CO line observations of T Tauri stars in the
Taurus-Auriga cloud clearly demonstrate that many low-mass
Pre-Main-Sequence (PMS) stars are surrounded by large ($R_{out}
\sim 200-800$~AU) Keplerian disks \citep{Koerner_93}, \citep{Dutrey_94}. There is little comparable evidence for such disks around
intermediate-mass PMS objects, the Herbig Ae stars.   One example of a molecular disk around an A-type star is  MWC480 (spectral type A4) which possesses a large CO disk \citep{Mannings_1997}. Near-infrared observations reveal 
that Herbig Ae stars can be surrounded by large reflection nebulae
as in the case of AB Auriga \citep{Grady_1999}, while ISO data
provide very strong  evidence for disk geometries
(e.g. AB Aur and HD 163296 \citep{Bouwman_2000}).
  However, to date, there
is little known about the molecular content and dynamics of these systems.

In this context, the A0 star HD 34282 ($\alpha = $05:16:00.47,
$\delta = - $09:48:35.3, J2000.0) appears a very interesting object
because i) the star has a strong IR excess \citep{JCMT, Malfait}, ii) it is a nearby star, according to Hipparcos measurements D$=160^{+60}_{-40}$pc, \citep{VdA} and iii) single-dish observations performed with the IRAM 30 m telescope (in November 1998) revealed a CO~\jdu~ double peak profile, strongly suggestive of a rotating disk, also observed by \citet{Greaves} in the CO~\jtd~ JCMT spectrum.
Moreover, \citet{Malfait} reported optical variability of the order of 2.5 mag in the V
band, which suggests that HD 34282 might be an UX Orionis-type star.

Interferometric observations of CO rotational lines remain the
best tool to search for large cold Keplerian disks similar
to those found around TTauri stars, while millimeter continuum emission provides complementary
information on the dust. We used the IRAM interferometer (PdBI) \footnote{Based on observations carried out with the IRAM Plateau de Bure Interferometer. IRAM is supported by INSU/CNRS (France), MPG (Germany) and
IGN (Spain)} to map the circumstellar material around HD 34282  simultaneously in \dco~\jdu~ at 1.3 mm and in continuum at 3.4 and 1.3 mm.




\section{Observational data}

The CO~\jdu~ and the continuum emissions from HD 34282 were mapped
with the PdBI, using three configurations: 4D1, in August 1999,
4B1 and 4C2+N09 in October 1999. Due to the low elevation of the
source, the resulting beam is strongly elliptical
($2.31 \arcsec \times 1.42 \arcsec $ at $\mathrm{PA} = 20 \degr$). The
spectral resolution was 0.10 km.s$^{-1}$ per channel at the CO
line frequency. We used full correlator power to look
simultaneously for dust continuum emission at 1.3 and 3.4 mm. Data
were reduced using the GILDAS package at IRAM Grenoble. At 1.3 mm, the seeing was about 0.3 \arcsec. Dirty maps were deconvolved using the classical CLEAN
algorithm. The integrated flux derived from the CO~\jdu~ spectrum observed at the IRAM 30m telescope (Pico Veleta,
Spain) in November 1998 is 4.8 $\pm$ 0.4 Jy.km.s$^{-1}$, while
the interferometric map leads to an integrated flux of 4.2 $\pm$ 0.1 
Jy.km.s$^{-1}$. As the two fluxes are in good agreement, we
are confident in the interferometric flux calibration, and we
conclude that our interferometric observations do not miss much flux.


\section{Data analysis}

\label{sec:analysis}

\subsection{Millimeter SED}

\label{sub:cont}

Following \citet{Beckwith_90}, we write the dust absorption
coefficient as $ \kappa_d(\nu)= 0.1 (\nu/10^{12}
\mathrm{Hz})^\beta  \mathrm{cm}^2.\mathrm{g}^{-1}$. As the spectral
index derived from the fluxes reported in Table \ref{tab:cont_data} is  
($\alpha = 3.15 \pm 0.20 $), we conclude that thermal dust emission 
from the circumstellar material around HD 34282 is optically thin at millimeter 
wavelengths. The corresponding dust opacity spectral index, $\beta = 1.15 \pm
0.20$, is significantly smaller than the spectral index of 2 \citep{Beckwith_90} measured 
in the interstellar medium.
The same behaviour is observed in T Tauri's disks \citep{Dutrey_96}. This similarly suggests that grain growth is occuring in the disk.

\begin{table}[h]
\begin{center}
\caption{Continuum fluxes of HD 34282}\label{tab:cont_data}
\begin{tabular}{ccl}
\hline
\hline
$\lambda$ & $F_\lambda$                         & References \\
(mm)      & (mJy)                               &            \\
\hline                                        
1.1 & 183\hphantom{.8} $ \pm $ 17\hphantom{.3}  & \citep{JCMT}\\
1.3 & 110\hphantom{.8} $ \pm $ 10\hphantom{.3}  & This work $^{\mathrm{a}}$  \\
2.6 & \hphantom{1}23.8 $ \pm $ \hphantom{1}3.0  & \citep{OVRO}\\
3.4 & \hphantom{12}5.0 $ \pm $ \hphantom{1}0.3  & This work  \\
\hline
\end{tabular}
\end{center}
\begin{list}{}{}
\item[$^{\mathrm{a}}$] We fitted an apparent size of 
$1.74 \pm 0.07 \arcsec \times 0.89 \pm 0.06 \arcsec$ at 1.3 mm. The disk is unresolved at 3.4 mm.
\end{list}
\end{table}

\subsection{Dust and CO PdBI data modeling}

We present in this section an improvement of the method described in \citet{DM_Tau}. It is based on a $\chi^2$ minimization in the $uv$ plane of a standard disk model, as we describe below. In the improved method, we combine the $^{12}$CO~\jdu~ line and the 1.3 mm continuum analysis in an iterative way. Assuming the $^{12}$CO~\jdu~line is optically thick, and the continuum at 1.3 mm is mainly optically thin (see previous section),  the method described below enables us to derive both temperature and density distributions with some ``classical'' assumptions: the dust and the gas have the same temperature, the gas to dust ratio is 100, the dust opacity at $10^{12}$ Hz is 0.1 cm$^{2}$ g$^{-1}$ and its spectral index $\beta$ is the one derived above.

The disk is assumed to be in local hydrostatic equilibrium, and the physical conditions in the disk are assumed to follow local thermodynamic  equilibrium. The physical parameters are described as power laws with the following radial dependences: 
$n(r) = n_{100} \times (r/100\rm{AU})^{-s}$ for H$_2$ density, $T(r)=T_{100} \times (r/100\rm{AU})^{-q}$ for kinetic temperature and $v(r) = v_{100} \times (r/100\rm{AU})^{-v}$ for rotation velocity. The surface density, $\Sigma(r)= \Sigma_{100} \times (r/100\rm{AU})^{-p}$, can be deduced from the scale height $H(r) = H_{100} \times (r/100\rm{AU})^{+h}$ and the density law $n(r)$ via $n(r) = {\Sigma(r)}/{\sqrt{\pi}H(r)}$.

\begin{figure}[h] {
{\includegraphics[width=3.7cm,angle=-90]{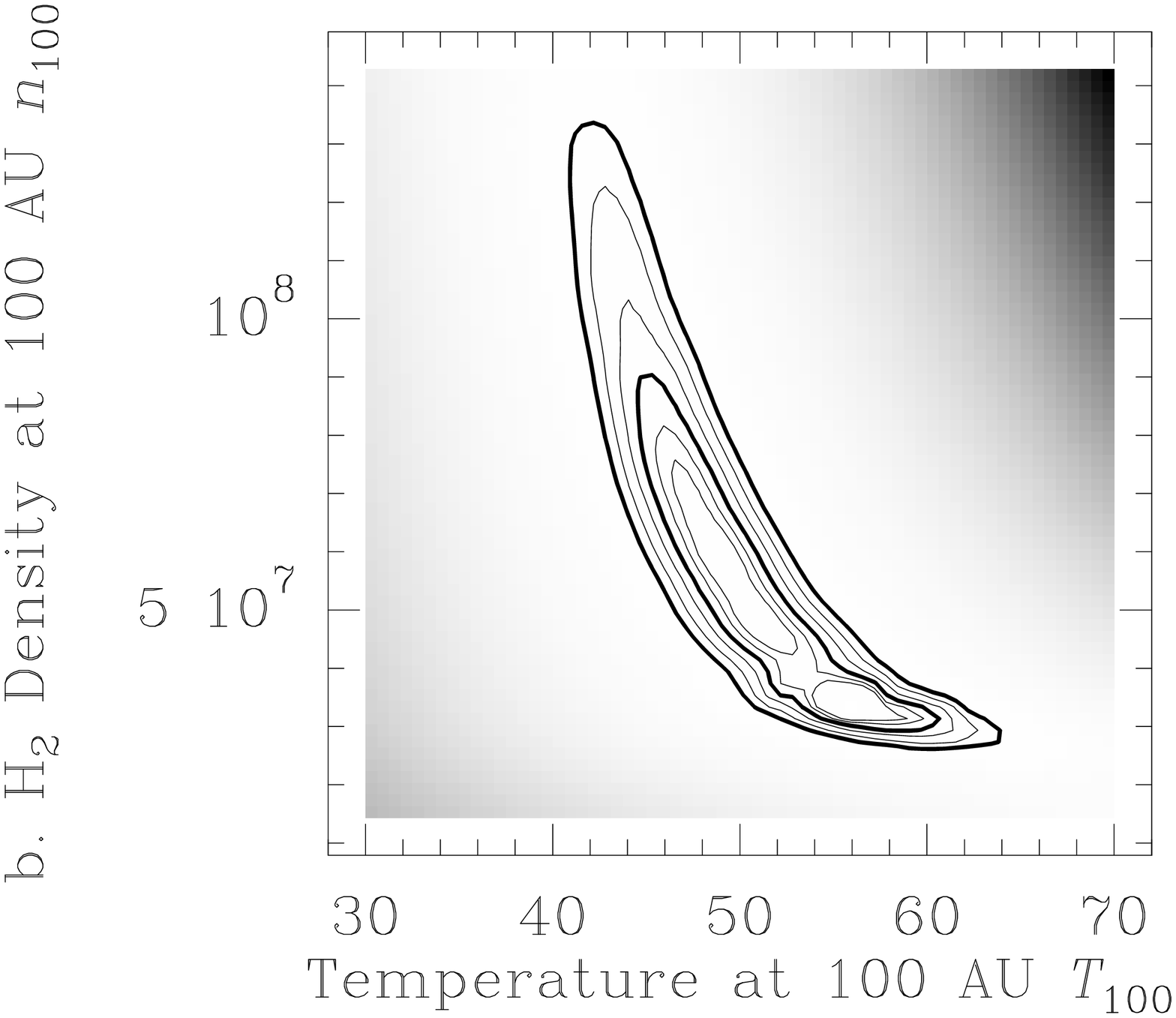} 
\includegraphics[width=3.7cm,angle=-90]{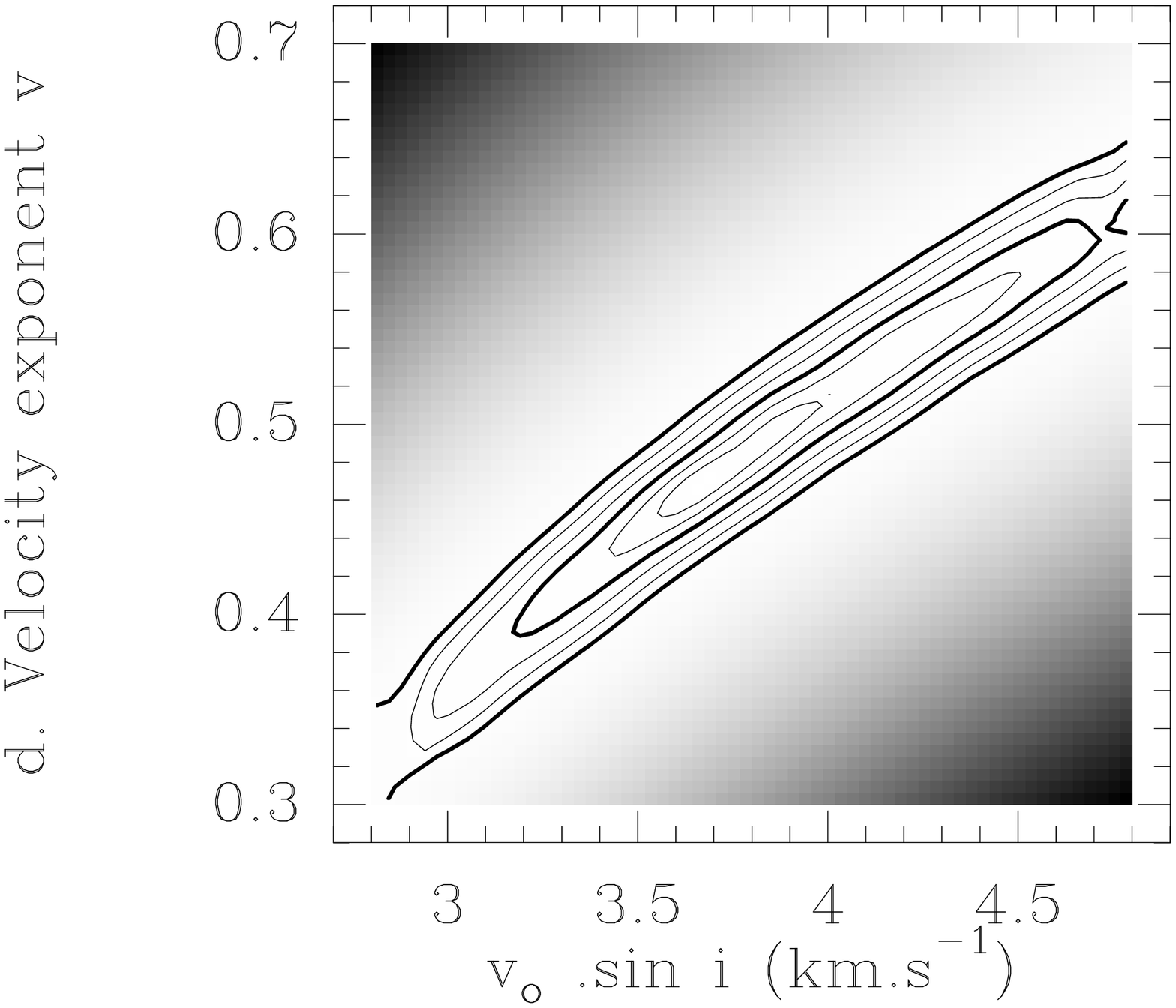}}
\vfill
{\includegraphics[width=3.65cm,angle=-90]{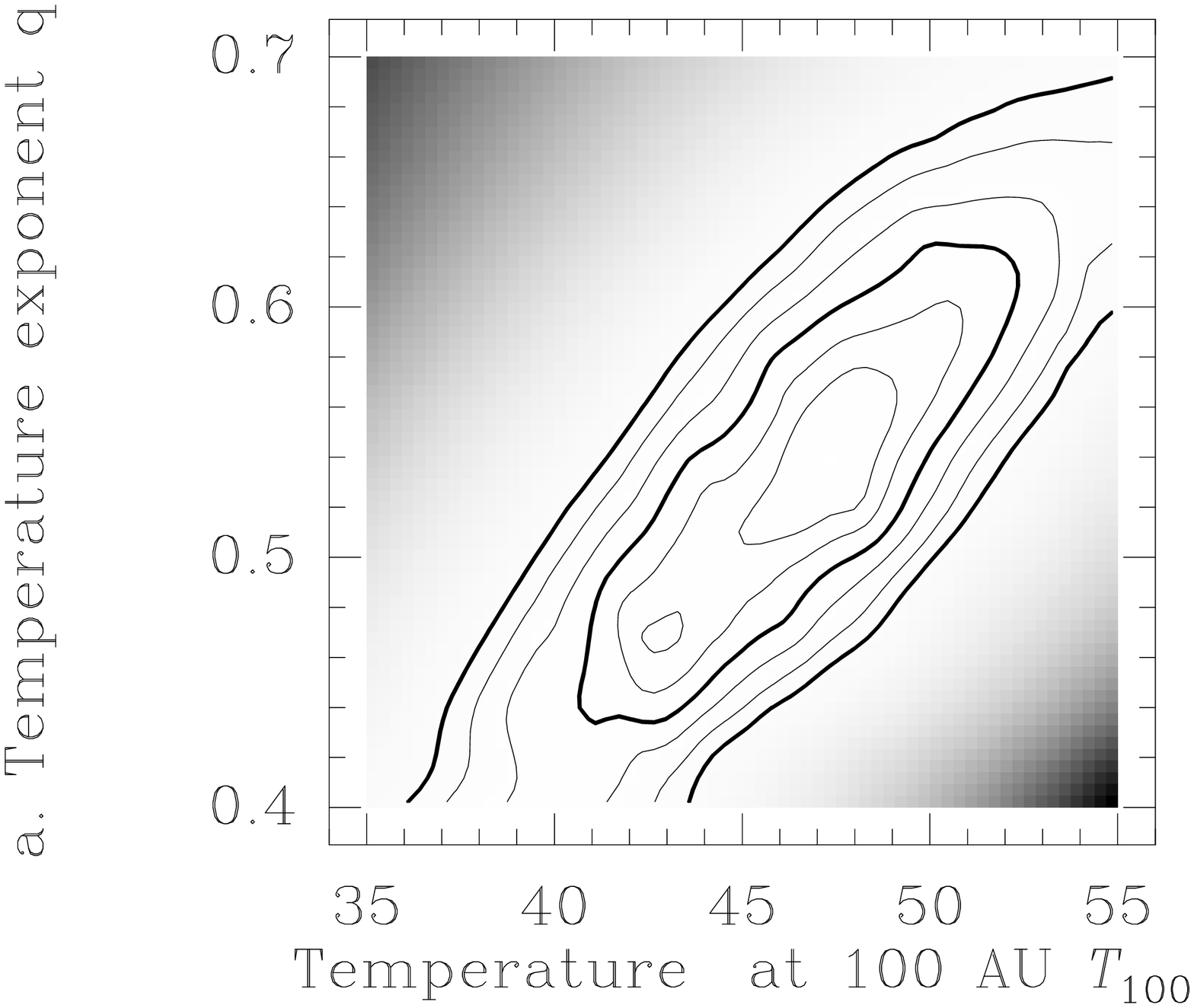} 
\includegraphics[width=3.7cm,angle=-90]{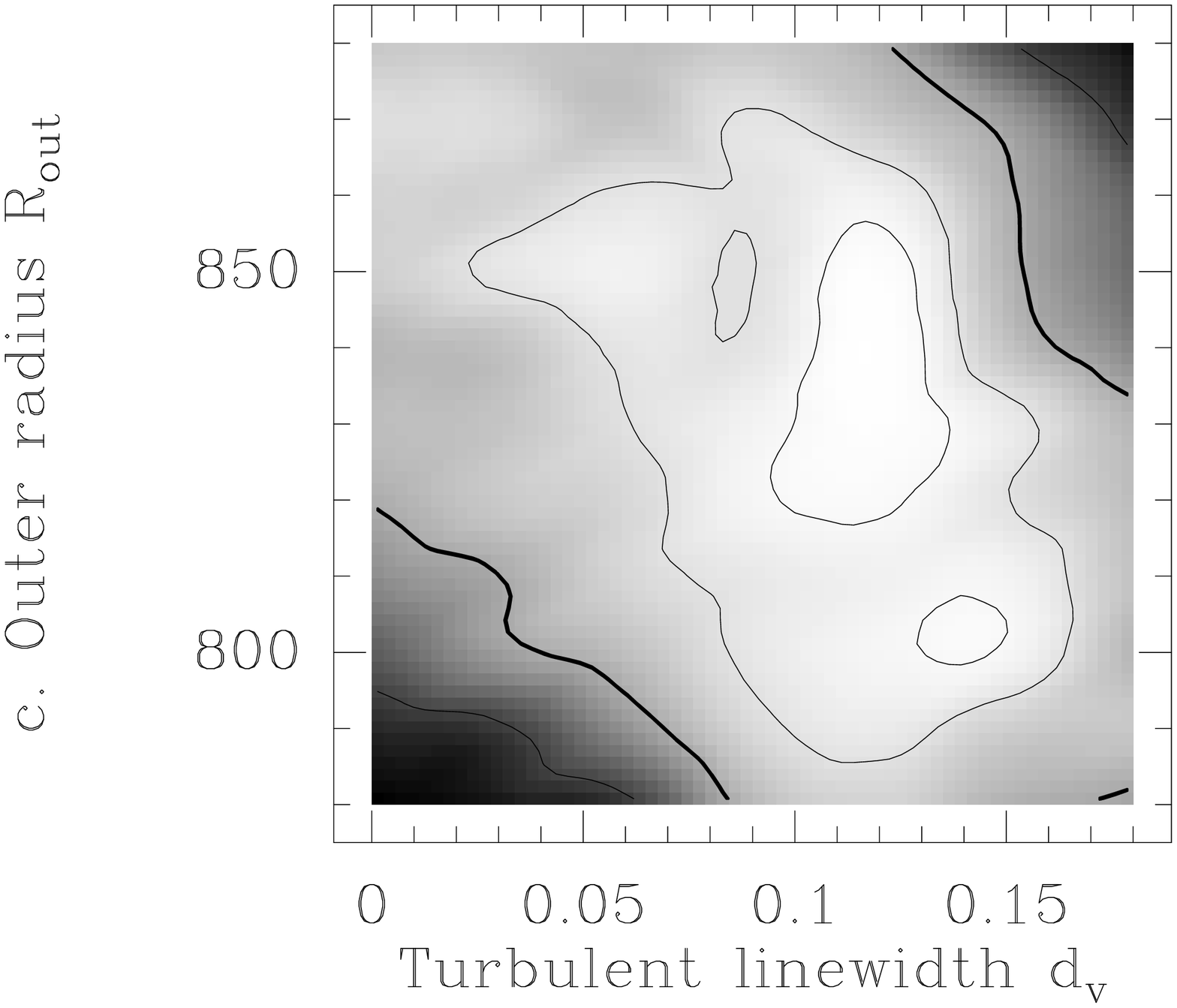}}
} \caption{Examples of the CO iso-$\chi^2$ surfaces. The contours
run from 1 to 6 $\sigma$. {\bf Top panels}: ($T_{100}$,$n_{100}$) and ($V_{100} \sin
i$,$v$) are coming from the $2 \times 2$ minimization. {\bf Bottom
panels}: ($T_{100}$,$q$) and ($\Delta v$,$R_{out}$) are from the final $4
\times 4$ minimization assuming the rotation pattern is Keplerian
($v=0.5$) and the density $n_{100}(\mathrm{dust})$ (see text).} \label{chi2_CO}
\end{figure}

For optically thick $^{12}$CO~\jdu~ emission, we cannot measure the
density distribution, but we are able to constrain the disk geometry (the outer radius $R_{out}$ and inclination $i$), the temperature profile and the turbulent velocity $\Delta v$ in the disk. Assuming a distance $D$, we first perform $\chi^2$  minimizations in the 2-dimensional parameter spaces (PA, V$_{lsr}$), (V$_{100}$, $v$), (V$_{100} \sin i$, $i$), ($T_{100}$, $q$) and ($\Delta v$, $R_{out}$) and then in the 5-dimensional and 4-dimensional spaces 
($T_{100},q,R_{out},i$ and $V_{100} \sin i $) and ($T_{100},q,R_{out},\Delta v$), which correspond to the more coupled parameters. 
The most important results of the minimizations are presented in
Fig.\ref{chi2_CO}. They will be discussed in next section.

A $\chi^2$ minimization performed on the parameters ($n_{100},T_{100}$)
shows that the line is optically thick ; above a threshold called $n_{100}(\rm{CO})$ in the following, the temperature becomes almost independent of the density. According to Fig.\ref{chi2_CO}-b, $n_{100}(\rm{CO})$ is close to the mimimum of the $\chi^2$ map ($n_{100}(\rm{CO})$, $T_{100}(\rm{CO})$) (note that this H$_2$ density 
assumes a standard CO abundance X$(\rm{CO})$ = X$^{12}_{\mathrm{TMC1}} = 7\times 10^{-5}$).

\begin{figure}[h]
{ {\includegraphics[width=3.6cm,angle=-90]{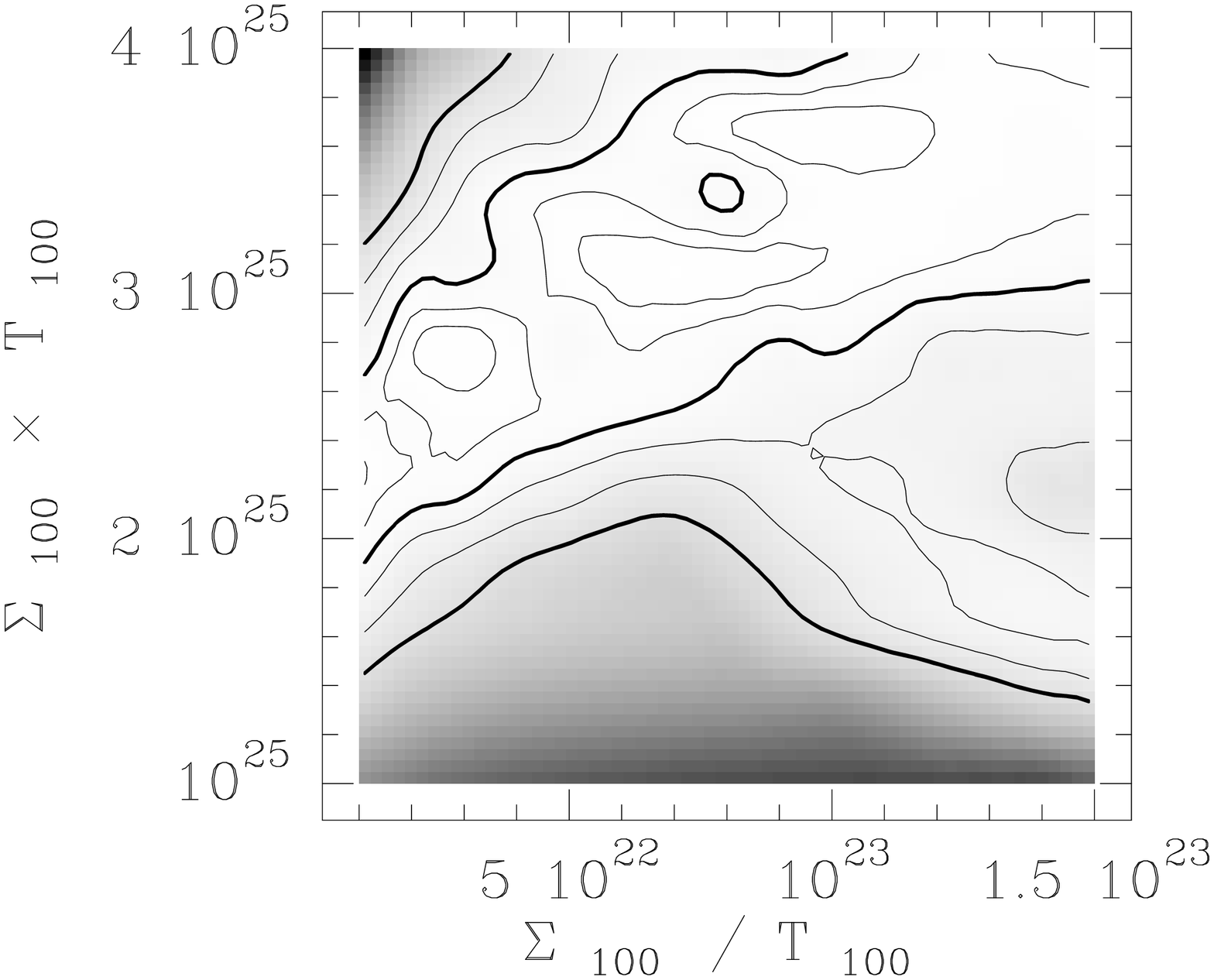} \hfill
\includegraphics[width=3.6cm,angle=-90]{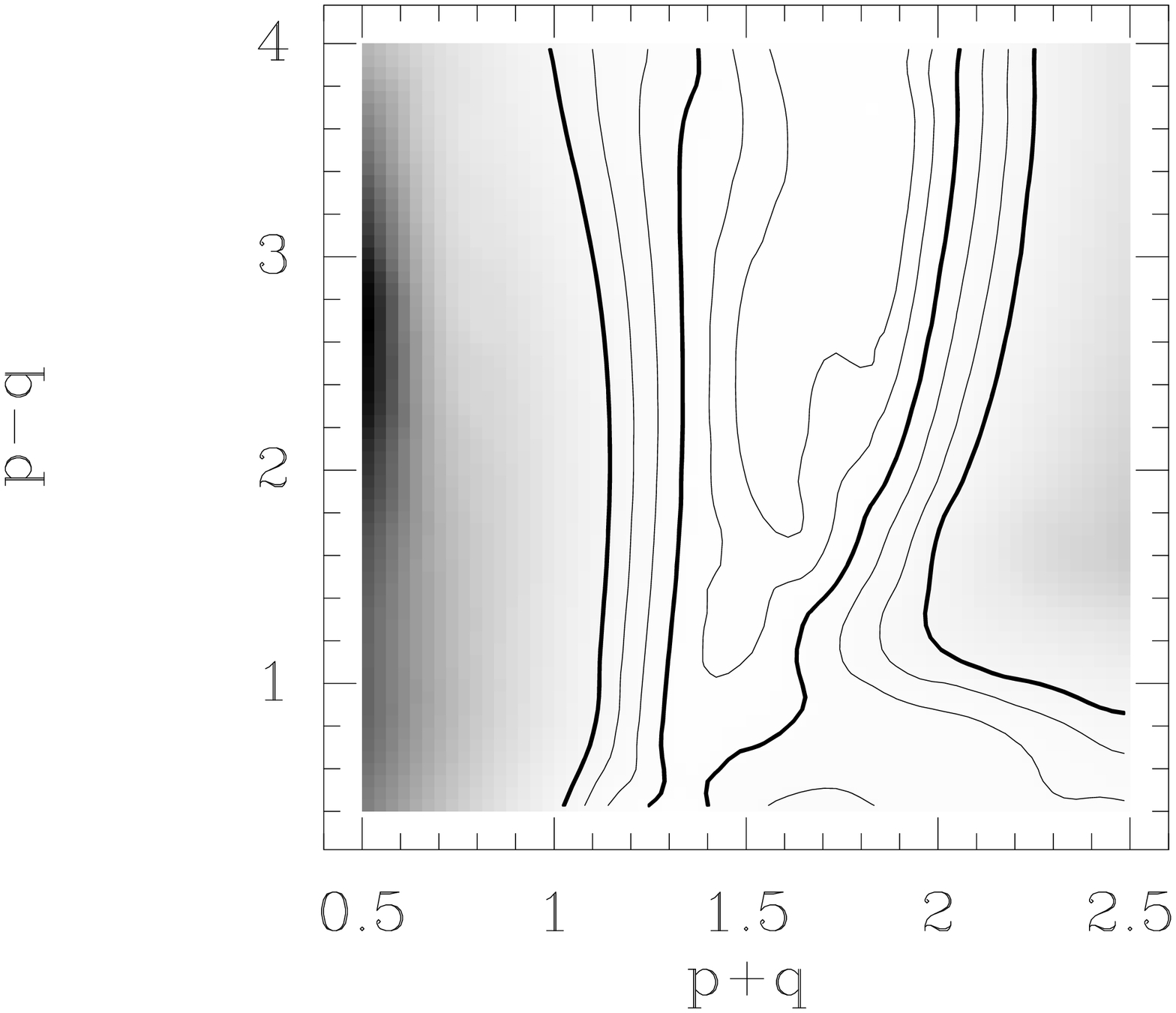}} }
\caption{Dust iso-$\chi^2$ surface. Contours run from 1 to 6 $\sigma$. {\bf Left}: ($\Sigma _{100} \times T_{100}$,
$\Sigma _{100} / T_{100}$); {\bf Right}: ($p+q$, $p-q$).} \label{cont}
\end{figure}

As the disk dust thermal emission is at least marginally
resolved, we may obtain an {\it independent estimate of the surface
density} (and the density, since the geometry is constrained) 
from the optically thin continuum map at 1.3mm.
Optically thin dust emissivity is
$\propto T_{100}\times \Sigma_{100} \times r^{-(p+q)}$.
Following the DM Tau disk analysis by Dartois \etal (in prep.) we
have performed a minimization on the parameters ($p+q$, 
$\Sigma _{100} \times T_{100}$, $p-q$, $\Sigma _{100} /T_{100}$). The results, 
given in Fig. \ref{cont}, confirm that dust emission is optically thin 
(since $p+q$ and $\Sigma_{100} \times T_{100} $ are relatively well constrained). 
Assuming the dust temperature law derived from \dco, we thus can derive $p$~and $\Sigma_{100}$ (or $s$ and $n_{100}$).

Comparison of the H$_2$ density lower limit, derived from the \dco~
 with the H$_2$ density derived from dust emission $n_{100}(\mathrm{dust})$, provides an lower limit to the CO abundance: $X(\mathrm{CO}) \geq n_{100}(\mathrm{CO})/n_{100}(\mathrm{dust}) \times \mathrm{X}_{\mathrm{TMC1}}^{12}\sim10^{-6}$. Equivalently, it also corresponds to an upper limit to the depletion factor $f(\dco)$ with respect to X$^{12}_{\mathrm{TMC1}}$: $f(\dco) \leq n_{100}(\mathrm{dust})/n_{100}(\mathrm{CO}) \sim 70$

\section{Discussion}

\subsection{Stellar Mass}

We have checked that the rotation pattern is almost Keplerian, with an
inferred value of the rotation exponent $v = 0.48 \pm 0.03 $ as shown
in Figure \ref{chi2_CO}-d. (We will then assume $v=0.5$ in the final modelling).

\label{sub:mass}

\citet{SDG_2000} have shown that, in such a case, the dynamical mass
(derived from CO modeling) provides an accurate measurement of the mass of the central star which scales as the distance $D$. 
Assuming a distance  $D = 160$~pc, we derive a star mass $M_{160} = 0.87 \pm 0.17$ M$_{\sun}$.

\subsection{Where is HD 34282 ?}

\begin{figure}[h]
\includegraphics[width=8.8 cm]{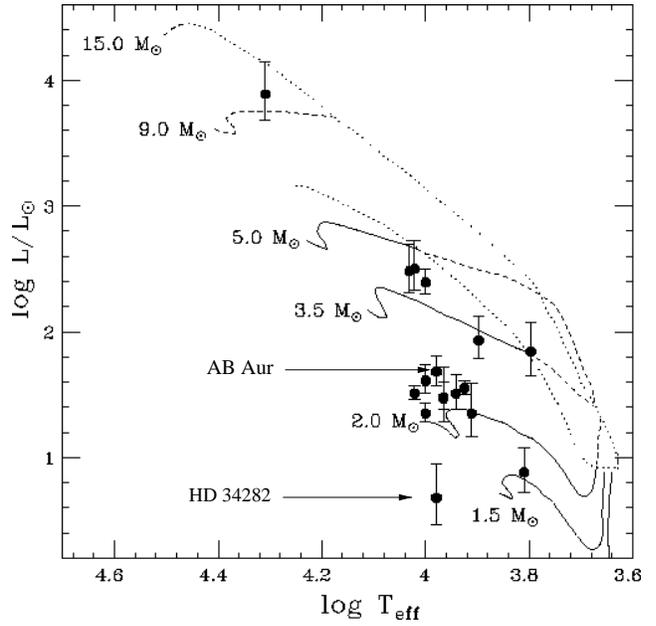}
\caption{Location of HD 34282 in a HR diagram for a distance $D=160$ pc (Fig. 1 of \citet{VdA})}
\label{ancker}
\end{figure}

This low mass value is unexpected for an A0 type star.
In addition, the luminosity derived from the photometric
measurements of \citet{VdA} is $L_{160}$ = 4.7 L$_{\sun}$, 
for a distance of 160 pc.  The implied location of HD 34282 in the H-R diagram places it at a
 totally different position to that of similar stars, such as
AB Aur, and is incompatible with stellar evolution tracks (see Figure
\ref{ancker}). The spectral type determination by \citet{Cannon} in the Henry Draper Catalog had been questioned by \citet{VdA}, but was subsequently confirmed by \citet{Gray}, who found the HD 34282 spectral type to be A0.5. Both ``anomalies'' point toward a revision of the
star distance.

As the luminosity, $L$, scales as $D^2$, and the dynamical mass $M$ 
as $D$, plotting HD 34282 in a
``distance-independent'' evolution diagram ($L/M^2$ vs
spectral type) allows a direct comparison with theoretical evolution
tracks (see Fig. 4). The error bars on the luminosity derived by 
\citet{VdA} ($\log L_{160} = 0.68 ^{+0.27}_{-0.21}$) are dominated by the distance
uncertainty. Assuming an arbitrary 10 \% uncertainty on the luminosity, we derive the  error bars for $L/M^2$ plotted in Fig. 4.

\begin{figure}[h]
\includegraphics[width=8.8 cm]{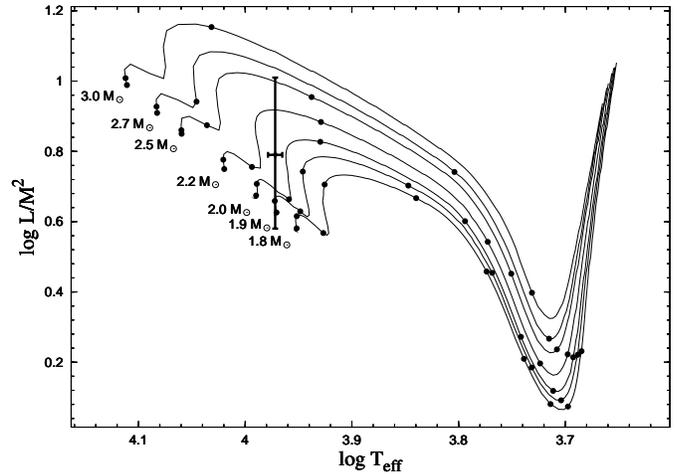}
\caption{Distance-independent diagram. Evolution tracks are from \citet{SDF2000} (available at \texttt{http://www-laog.obs.ujf-grenoble.fr/\-activites/\-star\-evol\-/evol.html}); dots are separated by 1 million years. Spectral-type A0.5 was converted to photospheric effective temperature using the results of \citet{Kenyon_hartmann} and an error bar of half a subclass was added in each direction} \label{fig_tracks}
\end{figure}

In this diagram, the star location corresponds to a stellar mass 
$M_* = 2.1^{+0.4} _{-0.2} M_{\sun}$, in better agreement with its A0 spectral type. From the dynamical mass M$_{160}$, we derive that actual distance of the star is $D = 160 \times (M_*/M_{160}) = 400^{+170} _{-100} $ pc. The distance derived from Hipparcos measurements on parallax ($6.1 \pm 1.6$ mas, \citet{VdA}) is more than $2 \sigma$ from our distance determination. From reddening considerations, \citet{JCMT} derived a distance of 547 pc, somewhat higher but still consistent with our determination. It should be noticed that considering its position, its distance, and its systemic velocity, HD 34282 is probably associated with Orion A \citep{H1_orion}.

\subsection{Distance revisited model}

At a distance $D$ = 400~pc, the $\chi^2$ analysis leads to the 
best parameters given in Table \ref{tab:1}. 

The disk outer radius $R_{out}$ is 835~AU, among the largest known disks 
(e.g. DM Tau, GG Tau).

\begin{table}[!ht]
\caption{Best Parameters for the HD 34282 disk (CO~\jdu~data and continuum).}
\label{tab:1}
\begin{tabular}{|l|rl|l|}
\hline
\hline
Assumed~ Distance & $D$ (pc)~ = & 400 & \\
\hline
Systemic velocity & $V_\mathrm{LSR}$ (km.s$^{-1}$)~= & --~2.35 & $\pm$ 0.02 \\
Orientation & PA~ = & 25 & $\pm$ 2$^{\circ} $ \\
Inclination & $i$~ = & 56 & $\pm$ 3$^{\circ} $ \\
\hline
Outer radius & $R_{out}$ (AU)~ = & 835 & $\pm$ $20$ \\
Turbulent width  & $\Delta v$ (km.s$^{-1}$)~ = & $0.11$ & $\pm$ $0.03$ \\
\hline
\multicolumn{4}{|c|}{Abundance \& H$_2$ Density law:~~ $n(r) = n_{100} (\frac{r}{100 \rm{AU}})^{- s}$} \\
\dco reference$^{(a)}$ &  X$^{12}_\mathrm{TMC1}$~ = & $7\times10^{-5}$ & - \\
\dco abundance &  X(\dco)~$>$ & $1\times10^{-6}$ &  - \\
\dco depletion &   $f$(\dco)~$<$         & $ 70$  & - \\
& Density& & \\
~~~~ at 100 AU& $n_{100}$ (cm$^{-3}$)~ = & $1.8\times10^9$ & $\pm 0.6$\\
~~~~ exponent       & $s$~=  & $2.45$ &$\pm 0.35$ \\
\hline
\multicolumn{4}{|c|}{Temperature law:~~~ $T(r)~ = T_{100} (\frac{r}{100\,\rm{AU}})^{-q}$ } \\
Temperature   &  &  &  \\
~~~~ at 100 AU  & $T_{100}$ (K)~ = & 46 & $\pm$ 5 \\
~~~~ exponent    & $q$~= & 0.52 & $\pm$ 0.08 \\
\hline

\multicolumn{4}{|c|}{Velocity law:~~~~~~ $V(r) = V_{100} (\frac{r}{100\,\rm{AU}})^{-v}$}\\
Velocity & & & \\
~~~~ at 100 AU    & $V_{100}$ (km.s$^{-1}$)~= & $4.6$& $\pm$ 0.2 \\
~~~~ exponent     & $v$~=        & $0.48$& $\pm$  0.03 \\
Stellar mass & M$_*$ ($\msun$)~= & 2.35 & $\pm$ 0.2 \\
\hline \hline
\multicolumn{4}{|c|}{Surface Density law:~~ $\Sigma(r) = \Sigma_{100} (\frac{r}{100\,\rm{AU}})^{- p}$} \\
Surface Density & & & \\
~~~~ at 100 AU  & $\Sigma_{100}$~ (cm$^{-2}$)~ = & $6\times10^{23}$ & $\pm$ 2.2 \\
& $\Sigma_{100}$~(g.cm$^{-2}$)~=&2.3 & $\pm$ 0.8\\
~~~~ exponent       & $p$~=&1.2 & $\pm$ 0.3 \\
\hline
\multicolumn{4}{|c|}{Scale Height law:~~ $H(r) = H_{100} (\frac{r}{100\,\rm{AU}})^{+ h}$} \\
Scale Height & & & \\
~~~~ at 100 AU      & $H_{100}$~(AU)~ = &13 & -\\
~~~~ exponent       & $h$~=  &1.24 & - \\
\hline \hline
\multicolumn{4}{|c|}{Dust:~~$\kappa_\nu = \kappa_{o}\times(\frac{\nu}
{10^{12}\,\mathrm{Hz}})^{\beta}$} \\
Absorption law & $\kappa_{o}$~=& $0.1$ & - \\
Dust exponent          & $\beta$~= & $1.15 $&$\pm 0.20$ \\
Dust disk size & $R_d$ (AU)~$>$ & 700 & - \\
~~~~~~ total mass & $M_d$ ($\msun$)~$=$ & $0.11^{+0.09}_{-0.05}$ & $$ \\ \hline
\end{tabular}

The errors are the 1$\sigma$ formal errors from the $\chi^2$ fit, and do not take into account the distance uncertainty.\\ $^{(a)}$
X$_\mathrm{TMC1}$, the \dco~abundance in TMC\,1 is taken from \citet{Cernicharo}.  \\ 

\end{table}

\begin{figure}[h]
\includegraphics[width=8.8cm]{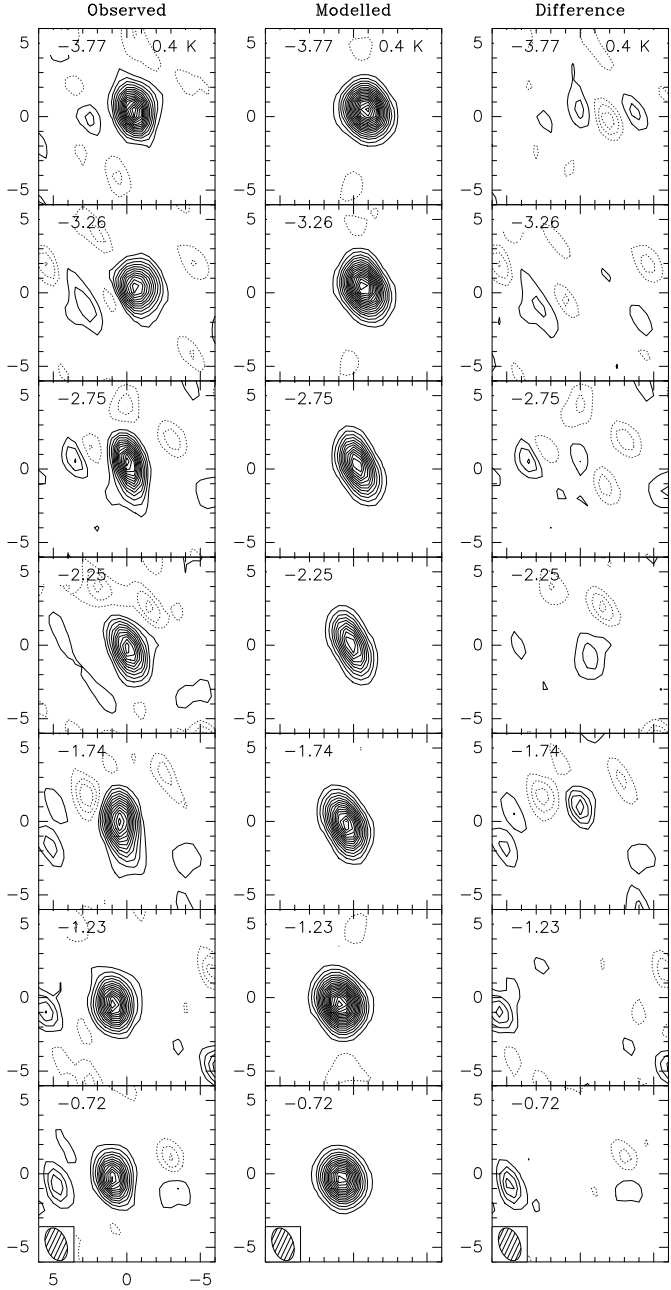}
\caption{{\bf Left}: \dco~\jdu observations. {\bf Middle}: Best model. {\bf Right}: Difference. Contours represents 50 mJy/beam, corresponding to a temperature of 0.35 K, or 1.5 $\sigma$. The channel velocity from top to bottom is: -3.77, -3.26, -2.75, -2.25. -1.74, -1.23 -0.72 km.s$^{-1}$}

\label{reg}
\end{figure}

It should be noticed that both the disk inclination angle of $i=56 ^\circ \pm 3
^\circ$ ($0 ^\circ$ means pole-on) and the measured photometric variability 
($\Delta V \simeq 2.5$) of HD 34282 are compatible with the UX Ori model 
from \citet{UXori}.

At 100~AU, the disk has $T_{100} = 46$~K which is hotter than found for T Tauri stars such as DM Tau or GM Aur \citep{Dutrey_98}, which  have similar radial variations of the temperature ($q=0.52 \pm 0.08$). Assuming the revised luminosity of 29 $\lsun$, the blackbody temperature at 100~AU would be around
50~K. The temperature law of the disk is therefore compatible with
a flared disk heated by the central star, e.g. \citep{Chiang_Goldreich}.

Assuming $T_{100} = 46$~K and $q=0.52$, we derive the surface
density $\Sigma_{100} \simeq 6.5 \times 10^{23}$cm$^{-2}$ and $p\simeq 1.2$
from the dust analysis (see Fig.\ref{cont}). With $p\simeq
1.2$, the surface density law is relatively shallow, as found in
the previous continuum analysis of TTauri disks by \citet{Dutrey_96}.

The disk mass derived from modeling of the continuum emission is 
$M_d = 0.11 ^{+0.09}_{-0.05} \msun$ (for a gas-to-dust
ratio of 100). Such a massive disk is in agreement with previous
results which suggest that there is a tendency for
intermediate-mass stars to have more massive disks \citep{Natta_2000}.

\section{Summary}

The disk surrounding the A0 PMS star HD 34282 has been mapped by
the IRAM array in $^{12}$CO~\jdu~and in continuum at 3.4 and
1.3 mm. Standard disk modeling of the molecular emission, 
combined with dust thermal emission analysis and stellar distance-independent 
evolution tracks, allow us to conclude that :

\begin{itemize}

\item The CO disk, resolved by the interferometric observations, is
in Keplerian rotation.

\item The disk outer radius is large, with $R_{out} = 835 \pm 20$~AU.

\item Its temperature law, derived from \dco~\jdu~ line is in agreement with a flared disk heated by the central star.

\item The disk continuum emission is optically thin and has a
spectral index $\beta \simeq 1.1$.

\item  With an inferred total mass of $0.11^{+0.09}_{-0.05} \msun$, the disk is relatively massive.

\item The stellar mass is $ 2.1 ^{+0.4}_{-0.2} \msun$ (as expected 
for a A0.5 spectral type star) and the star distance is $D =400^{+170} _{-100} $ pc, at a little bit more than $2 \sigma$ from the Hipparcos measurements. The revised luminosity would rather be around $L = 29 ^{+30}_{-13}~$L$_{\odot}$.

\end{itemize}

In conclusion, the HD 34282 disk does not appear significantly
different from the TTauri disks, except that it is more massive
and somewhat hotter, as expected for an intermediate mass-star.

\label{concl}

\begin{acknowledgements}

We warmly thank Dr. Mike Jura, who initiated this work during a
visit of Anne Dutrey at UCLA, for his careful reading of the
manuscript. Dimitri Pourbaix is acknowledged for providing
us with many useful comments about Hipparcos analysis. St\'ephane Guilloteau and Fr\'ed\'eric Gueth are thanked for fruitful discussions.
This research has made use of the SIMBAD database, operated at CDS, Strasbourg, France. We also would like to acknowledge the IRAM Plateau
de Bure staff for providing the observations which were performed
in service observing.

\end{acknowledgements}


\bibliography{BIBFILES/lettre,tata}

\bibliographystyle{aa}


\end{document}